\documentclass[12pt,a4paper]{article}
\usepackage{epsfig}
\newcommand{\be}{\begin{equation}}
\newcommand{\ee}{\end{equation}}
\newcommand{\bea}{\begin{eqnarray}}
\newcommand{\eea}{\end{eqnarray}}

\begin{document}
\begin{center}
{\hbox to\hsize{\hfill KEK-TH-961}}

\vspace{4\baselineskip}

\textbf{\Large Detailed Analysis of Proton Decay Rate \\
in the Minimal Supersymmetric SO(10) Model}\par
\bigskip
\vspace{2\baselineskip}
\textbf{\large Takeshi Fukuyama$^{\dagger}$
\footnote{E-Mail: fukuyama@se.ritsumei.ac.jp}},  
\textbf{\large Amon Ilakovac$^{\ddagger}$
\footnote{E-Mail: ailakov@rosalind.phy.hr}},  
\textbf{\large Tatsuru Kikuchi$^{\dagger}$
\footnote{E-Mail: rp009979@se.ritsumei.ac.jp}}, \\  
\textbf{\large Stjepan Meljanac$^{\star}$
\footnote{E-Mail: meljanac@irb.hr}}  
\textbf{\large and Nobuchika Okada$^{\diamond}$
\footnote{E-Mail: nobuchika.okada@kek.jp}} \\
\vspace{1\baselineskip}
\medskip{\it
$^\dagger$Department of Physics, Ritsumeikan University, 
Kusatsu, Shiga, 525-8577 Japan\\
$^\ddagger$University of Zagreb,
Department of Physics,
P.O. Box 331,
Bijeni\v cka cesta 32,
HR-10002 Zagreb, Croatia\\
$^\star$Institut Rudjer Bo\v skovi\'c,
Bijeni\v cka cesta 54,
P.O. Box 180,
HR-10002 Zagreb, Croatia\\
$^\diamond$Theory Division, KEK, Oho 1-1, Tsukuba, 
Ibaraki 305-0801, Japan}\\
%
%
\end{center}
\vspace{1\baselineskip}
\begin{abstract}
We consider the minimal supersymmetric $SO(10)$ model, 
 where only one {\bf 10} and one $\overline{\bf 126}$ Higgs multiplets 
 have Yukawa couplings with matter multiplets. 
This model has the high predictive power 
 for the Yukawa coupling matrices 
 consistent with the experimental data of 
 the charged fermion mass matrices, 
 and all the Yukawa coupling matrices are completely determined 
 once a few parameters in the model are fixed. 
This feature is essential for definite predictions 
 to the proton decay rate through the dimension five operators. 
Althogh it is not completely general, 
we analyze the proton decay rate for the dominant 
 decay modes $p \to K^{+} \overline{\nu}$ 
 by including as many free parameters as possible and varying them. 
There are two free parameters in the Yukawa sector, 
 while three in the Higgsino sector. 
It is found that an allowed region exists 
 when the free parameters in the Higgs sector 
 are tuned so as to cancel the proton decay amplitude. 
The resultant proton lifetime is proportional to 
 $1/\tan^2 \beta$ 
 and the allowed region eventually disappears 
 as $\tan \beta$ becomes large. 
\end{abstract}
\bigskip
\vspace{2\baselineskip}
\newpage
\section{Introduction}

One particularly attractive idea for the physics beyond 
 the standard model (SM) is the possible appearance of 
 a grand unified theory (GUT) \cite{gut}, 
 where the standard model gauge interactions are 
 all unified into a simple gauge group. 
The successful gauge coupling unification 
 of the minimal supersymmetric (SUSY) standard model (MSSM) 
 at a scale $M_{G} \simeq 2 \times 10^{16}\, \mathrm{[GeV]}$ 
 \cite{susygut2}  
 strongly suggests the circumstantial evidence 
 of both ideas of SUSY and GUT, namely the SUSY GUTs. 

The most characteristic prediction of the (SUSY) GUTs 
 is the proton decay. 
Normally in SUSY GUTs the proton decay process 
 through the dimension five operators 
 involving MSSM matters and the color triplet Higgsino 
 turns out to be the dominant decay modes, 
 since the process is suppressed by only a power of 
 the Higgsino mass scale. 
Experimental lower bound on the proton decay modes 
  $p \to K^{+} \overline{\nu}$ through the dimension five operators 
  is given by SuperKamiokande (SuperK) \cite{sk}, 
\be
\tau (p \rightarrow K^{+} \overline{\nu}) \geq 
 2.2 \times 10^{33} \,\, [{\mathrm{years}}].  
\ee
This is one of the most stringent constraints 
 in construction of realistic SUSY GUT models.
In fact, the minimal SUSY $SU(5)$ model has been argued 
 to be excluded from the experimental bound 
 together with the requirement 
 of the success of the three gauge coupling unification 
 \cite{mp} \cite{gn}. 
However note that 
 the minimal $SU(5)$ model predictions contradicts 
 against the realistic charged fermion mass spectrum,  
 and thus, strictly speaking, the model is ruled out 
 from the beginning. 
Obviously some extensions of the flavor structures in the model 
 is necessary to accommodate the realistic fermion mass spectrum. 
On the other hand, knowledge of the flavor structure is essential 
 in order to give definite predictions about 
 the proton decay processes through the dimension five operators. 
Some models in which flavor structures are extended 
 have been found to be consistent with the experiments 
 \cite{bps} \cite{ew}.  

Recently, masses and flavor mixings of neutrinos 
 have been confirmed through the neutrino oscillation 
 phenomena. 
This is the evidence of new physics beyond the standard model. 
One of the most promising candidates for new physics 
 naturally accommodating the neutrino physics  
 is the SUSY GUTs based on the gauge group $SO(10)$. 
This is because in $SO(10)$ models 
 all the matters in the standard model 
 together with additional right-handed neutrinos 
 are unified into a single representation ${\bf 16}$. 
Also the models can naturally explain the tiny neutrino masses 
 compared to the electroweak scale 
 through the seesaw mechanism \cite{seesaw}. 
Lots of models based on $SO(10)$ have been proposed and 
 extensively discussed. 
To be realistic, any models must accommodate 
 all the experimental data of the fermion mass matrices 
 and also be consistent with the experimental lower bound 
 of the proton lifetime. 
As discussed above, knowledge of the flavor structure 
 in a  given model is necessary in order to give 
 a definite results on the proton lifetime. 

In this paper, we consider the so-called minimal SUSY $SO(10)$ model 
proposed in \cite{bm} and analyzed in detail in \cite{fo} \cite{gmn} 
\cite{mimura}, 
 and we perform detailed analysis on the proton decay rate. 
The important fact is that 
 this model has the high predictive power for the Yukawa matrices 
 with leaving a few parameters free. 
Fixing these parameters, we can completely determine 
 all the elements in the Yukawa matrices, whose information 
 is essential for the definite predictions to the proton decay 
 process through the dimension five operators. 
\footnote{The importance of flavor physics in dimension six proton decay 
processes is discussed in \cite{perez}. }
Information of the Higgs sector in the model is 
 also necessary for the definite predictions, 
 and we include it in our analysis as free parameters. 
For simplicity, we assume all the soft masses are flavor diagonal. 

This paper is organized as follows: 
in the next section, we consider the minimal SUSY $SO(10)$ model, 
 and show the fact that the model has 
 the high predictive power for the Yukawa matrices. 
In section 3 we analyze the proton decay processes 
 with some free parameters and the predicted Yukawa matrices 
 in the model. 
We show the distributions of the predicted proton lifetime 
for typical values of the free parameters. 
Section 4 is devoted to conclusions and discussions.

\section{Minimal SO(10) model and its predictions} 

We begin with a review of the minimal SUSY $SO(10)$ model
 proposed in \cite{bm} and recently analyzed in detail 
 in Ref. \cite{fo} \cite{gmn} \cite{mimura}. 
\footnote{
In this section, we repeat almost the same discussions in 
Ref. \cite{fo} 
(but input values we use are different) 
for the convenience of readers. 
If a reader is familiar with discussions here, 
the reader can skip to the the last paragraph of this section. } 
Even when we concentrate our discussion on the issue how to 
 reproduce the realistic fermion mass matrices in the $SO(10)$ model, 
there are lots of possibilities for introduction of Higgs multiplets.  
The minimal supersymmetric $SO(10)$ model is the one where only 
 one {\bf 10} and one $\overline{\bf 126}$ Higgs multiplets have 
 Yukawa couplings  (superpotential) with {\bf 16} matter multiplets 
 such as 
\begin{eqnarray}
W_Y = Y_{10}^{ij} {\bf 16}_i {\bf 10}_H {\bf 16}_j 
+ Y_{126}^{ij} {\bf 16}_i {\bf \overline{126}}_H {\bf 16}_j \; , 
\label{Yukawa1}
\end{eqnarray} 
where ${\bf 16}_i$ is the matter multiplet of the $i$-th generation, 
 ${\bf 10}_H$ and ${\bf \overline{126}}_H$ are the Higgs multiplet of 
 ${\bf 10}$ and $\overline{\bf 126}$ representations under $SO(10)$, 
 respectively.  
Note that, by virtue of the gauge symmetry, 
 the Yukawa couplings, $Y_{10}$ and $Y_{126}$, are complex symmetric 
 $3 \times 3$ matrices.  
We assume some appropriate Higgs multiplets, whose vacuum expectation 
 values (VEVs) correctly break the $SO(10)$ GUT gauge symmetry 
 into the standard model one at the GUT scale, 
 $M_{G} \simeq 2 \times 10^{16} \mathrm{[GeV]}$.  
Suppose the Pati-Salam subgroup \cite{ps}, 
 $G_{422}=SU(4)_c \times SU(2)_L \times SU(2)_R$, 
 at the intermediate breaking stage.  
Under this symmetry, the above Higgs multiplets are decomposed as 
 ${\bf 10} \rightarrow 
 ({\bf 6},{\bf 1},{\bf 1}) + ({\bf 1},{\bf 2},{\bf 2}) $ 
 and 
 $\overline{\bf 126} \rightarrow 
 ({\bf 6}, {\bf 1}, {\bf 1} ) 
 + ( {\bf \overline{10}}, {\bf 3}, {\bf 1}) 
 + ({\bf 10}, {\bf 1}, {\bf 3})  
 + ({\bf 15}, {\bf 2}, {\bf 2}) $, 
 while ${\bf 16} \rightarrow ({\bf 4}, {\bf 2}, {\bf 1}) 
 + (\overline{\bf 4}, {\bf 1}, {\bf 2})$.  
Breaking down to the standard model gauge group, 
 $SU(4)_c \times SU(2)_R  \rightarrow SU(3)_c \times U(1)_Y$,   
 is accomplished by non-zero VEV  of 
 the $({\bf 10}, {\bf 1}, {\bf 3})$ Higgs multiplet. 
Note that Majorana masses for the right-handed neutrinos 
 are also generated by this VEV through the Yukawa coupling 
 $Y_{126}$ in Eq.~(\ref{Yukawa1}).  
In general, the $SU(2)_L$ triplet Higgs in 
 $({\bf \overline{10}}, {\bf 3}, {\bf 1}) \subset \overline{\bf 126}$ 
 would obtain the VEV induced through the electroweak symmetry breaking 
 and may play a crucial role 
 of the light Majorana neutrino mass matrix. 
This model is called the type II seesaw model, 
 and we include this possibility in our model. 

After the symmetry breaking, we find two pair of Higgs doublets 
 in the same representation as the pair in the MSSM.  
One pair comes from $({\bf 1},{\bf 2},{\bf 2}) \subset {\bf 10}$ 
 and the other comes from 
 $({\bf 15}, {\bf 2}, {\bf 2}) \subset \overline{\bf 126}$.  
Using these two pairs of the Higgs doublets, 
 the Yukawa couplings of Eq.~(\ref{Yukawa1}) are rewritten as 
\begin{eqnarray}
W_Y &=& (U^c)_i  \left(
Y_{10}^{ij}  H^u_{10} + Y_{126}^{ij}  H^u_{126} \right) Q_j 
+ (D^c)_i  \left(
Y_{10}^{ij}  H^d_{10} + Y_{126}^{ij}  H^d_{126} \right) Q_j  
\nonumber \\ 
&+& (N^c)_i \left( 
Y_{10}^{ij}  H^u_{10} - 3 Y_{126}^{ij} H^u_{126} \right) L_j 
+ (E^c)_i  \left(
Y_{10}^{ij}  H^d_{10}  - 3 Y_{126}^{ij} H^d_{126} \right) L_j   
\nonumber \\
&+&
 L_i \left( Y_{126}^{ij} \; v_T \right) L_j +
(N^c)_i  
 \left( Y_{126}^{ij} \; v_R \right) 
(N^c)_j \;  , 
\label{Yukawa2}
\end{eqnarray} 
where $U^c$, $D^c$, $N^c$ and $E^c$ are the right-handed $SU(2)_L$ 
 singlet quark and lepton superfields, $Q$ and $L$ 
 are the left-handed $SU(2)_L$ doublet quark and lepton superfields, 
 $H_{10}^{u,d}$ and $H_{126}^{u,d}$ are up-type and down-type 
 Higgs doublet superfields originated 
 from ${\bf 10}$ and ${\bf \overline{126}}$, respectively, 
 and the last two terms is the Majorana mass term of 
 the left-handed and the right-handed neutrinos, respectively, 
 developed by the VEV of 
 the $({\bf \overline{10}}, {\bf 3}, {\bf 1})$ Higgs ($v_T$) 
 and the $({\bf 10}, {\bf 1}, {\bf 3})$ Higgs ($v_R$).  
The factor $-3$ in the lepton sector is the Clebsch-Gordan 
coefficient.  

In order to keep the successful gauge coupling unification, 
 suppose that one pair of Higgs doublets 
 given by a linear combination $H_{10}^{u,d}$ and $H_{126}^{u,d}$ 
 is light while the other pair is  heavy ($\simeq M_G$).  
The light Higgs doublets are identified as 
 the MSSM Higgs doublets ($H_u$ and $H_d$) and given by 
\begin{eqnarray} 
H_u &=& \widetilde{\alpha}_u  H_{10}^u 
+ \widetilde{\beta}_u  H_{126}^u, 
\nonumber \\
H_d &=& \widetilde{\alpha}_d  H_{10}^d  
+ \widetilde{\beta}_d  H_{126}^d  \; , 
\label{mix}
\end{eqnarray} 
where $\widetilde{\alpha}_{u,d}$ and $\widetilde{\beta}_{u,d}$ denote 
elements of the unitary matrix  which rotate the flavor basis in the 
original model into the (SUSY) mass eigenstates.  
Omitting the heavy Higgs mass eigenstates, the low energy 
superpotential is described by only the light Higgs doublets 
$H_u$ and $H_d$ such that 
\begin{eqnarray}
W_Y &=& 
(U^c) _i \left( \alpha^u  Y_{10}^{ij} + 
\beta^u  Y_{126}^{ij} \right)  H_u \, Q_j 
+ (D^c)_i  
\left( \alpha^d  Y_{10}^{ij} + 
\beta^d   Y_{126}^{ij}  \right) H_d \,Q_j  \nonumber \\ 
&+& (N^c)_i  
\left( \alpha^u  Y_{10}^{ij} -3 
\beta^u   Y_{126}^{ij} \right)  H_u \,L_j 
+ (E^c)_i  
\left( \alpha^d  Y_{10}^{ij} -3 
\beta^d   Y_{126}^{ij}  \right) H_d \,L_j \nonumber \\ 
&+& 
  L_i \left( Y_{126}^{ij} \; v_T \right) L_j + 
 (N^c)_i  
  \left( Y_{126}^{ij} v_R \right)  (N^c)_j \; ,  
\label{Yukawa3}
\end{eqnarray} 
where the formulas of the inverse unitary transformation 
 of Eq.~(\ref{mix}), 
 $H_{10}^{u,d} = \alpha^{u,d} H_{u,d} + \cdots $ and 
 $H_{126}^{u,d} = \beta^{u,d} H_{u,d} + \cdots $, have been used. 

Providing the Higgs VEVs, $H_u = v \sin \beta$ and $H_d = v \cos \beta$ 
with $v \simeq 174.1 \mbox{[GeV]}$, the quark and lepton mass matrices can be 
read off as
\begin{eqnarray}
 M_u &=& c_{10} M_{10} + c_{126} M_{126}, 
 \nonumber \\
 M_d &=& M_{10} + M_{126},   
 \nonumber \\
 M_D &=& c_{10} M_{10} - 3 c_{126} M_{126}, 
 \nonumber \\
 M_e &=& M_{10} - 3 M_{126}, 
 \nonumber \\
 M_T &=& c_T M_{126},
 \nonumber \\ 
 M_R &=& c_R M_{126}, 
\label{massmatrix}
\end{eqnarray} 
where $M_u$, $M_d$, $M_D$, $M_e$, $M_T$ and $M_R$ denote 
 up-type quark, down-type quark, 
 neutrino Dirac, charged-lepton, 
 left-handed neutrino Majorana, 
 and right-handed neutrino Majorana mass matrices, respectively.  
Note that all the quark and lepton mass matrices are characterized 
by only two basic mass matrices, $M_{10}$ and $M_{126}$, 
 and four complex coefficients 
 $c_{10}$, $c_{126}$, $c_T$ and $c_R$, which are defined as 
 $M_{10}= Y_{10}\, \alpha^d  v \cos\beta$, 
 $M_{126} = Y_{126}\, \beta^d  v \cos\beta$, 
 $c_{10}= (\alpha^u/\alpha^d) \tan \beta$, 
 $c_{126}= (\beta^u/\beta^d) \tan \beta $, 
 $c_T = v_T/( \beta^d  v  \cos \beta)$, and 
 $c_R = v_R/( \beta^d  v  \cos \beta)$, respectively. 

The mass matrix formulas in Eq.~(\ref{massmatrix}) 
 leads to the GUT relation 
 among the quark and lepton mass matrices, 
\begin{eqnarray}
M_e = c_d \left( M_d + \kappa  M_u \right) \; , 
\label{GUTrelation} 
\end{eqnarray} 
where 
\begin{eqnarray}
c_d &=& - \frac{3 c_{10} + c_{126}}{c_{10}-c_{126}}, 
\nonumber \\
\kappa &=& - \frac{4}{3 c_{10} + c_{126}}. 
\end{eqnarray} 
Without loss of generality, we can begin with the basis 
 where $M_u$ is real and diagonal, $M_u = D_u$.  
Since $M_d$ is the symmetric matrix, it can be described 
 as $M_d = V_{\mathrm{CKM}}^* \,D_d 
\, V_{\mathrm{CKM}}^\dagger$ 
 by using the CKM matrix $V_{\mathrm{CKM}}$ 
 and the real diagonal mass matrix $D_d$.  
\footnote{
In general, $M_d = \overline{V}_{\mathrm{CKM}}^{\,*} \,D_d 
\, \overline{V}_{\mathrm{CKM}}^{\,\dagger}$ by using 
 a general unitary matrix 
 $\overline{V}_{\mathrm{CKM}}=e^{i \alpha} e^{i \beta T_3} e^{i \gamma T_8} 
    V_{\mathrm{CKM}} 
    e^{i \beta^\prime T_3} e^{i \gamma^\prime T_8}$. 
We omit the diagonal phases to keep the free parameters in the model 
 minimum as possible. 
We can check that 
 the final results of our analysis for the proton lifetime 
 remains almost the same even if the diagonal phases are included. }
Considering the basis-independent quantities,  
 $\mathrm{tr} [M_e^\dagger M_e ]$, 
 $\mathrm{tr} [(M_e^\dagger M_e)^2 ]$ 
 and $\mathrm{det} [M_e^\dagger M_e ]$, 
 and eliminating $|c_d|$, we obtain two independent equations,  
\begin{eqnarray}
\left(
\frac{\mathrm{tr} [\widetilde{M_e}^\dagger \widetilde{M_e} ]}
{m_e^2 + m_{\mu}^2 + m_{\tau}^2} \right)^2
&=& 
\frac{\mathrm{tr} [( \widetilde{M_e}^\dagger \widetilde{M_e} )^2 ]}
{m_e^4 + m_{\mu}^4 + m_{\tau}^4},
\label{cond1} \\ 
\left( \frac{\mathrm{tr} [\widetilde{M_e}^\dagger \widetilde{M_e} ]}
{m_e^2 + m_{\mu}^2 + m_{\tau}^2} \right)^3
&=&
\frac{\mathrm{det} [\widetilde{M_e}^\dagger \widetilde{M_e} ]}
{m_e^2 \; m_\mu^2 \; m_\tau^2},
\label{cond2} 
\end{eqnarray}
where $\widetilde{M_e} 
 \equiv V_{\mathrm{CKM}}^* \, 
D_d \, V_{\mathrm{CKM}}^\dagger 
 + \kappa D_u$.  
With input data of six quark masses, 
 three angles and one CP-phase in the CKM matrix 
 and three charged lepton masses, 
 we can solve the above equations 
 and determine $\kappa$ and $|c_d|$, 
 but one parameter, the phase of $c_d$, 
is left undetermined \cite{fo}. 
The original basic mass matrices, $M_{10}$ and $M_{126}$, 
 are described by 
\begin{eqnarray}
M_{10} 
&=& 
\frac{3+ |c_d| e^{i \sigma}}{4} 
\overline{V}_{\mathrm{CKM}}^{\,*} 
\, D_d \, \overline{V}_{\mathrm{CKM}}^{\,\dagger}
+ \frac{|c_d| e^{i \sigma} \kappa}{4} D_u, 
\label{M10}  \\ 
M_{126} &=& 
 \frac{1- |c_d| e^{i \sigma}}{4} 
 \overline{V}_{\mathrm{CKM}}^{\,*} \, D_d \, 
\overline{V}_{\mathrm{CKM}}^{\,\dagger}
 -\frac{|c_d| e^{i \sigma} \kappa}{4} D_u. 
 \label{M126} 
\end{eqnarray} 
Here the unitary matrix $\overline{V}_{\mathrm{CKM}}$ 
is reltated with the conventional $V_{\mathrm{CKM}}$ as 
\be
\overline{V}_{\mathrm{CKM}}=e^{i \alpha} e^{i \beta T_3} e^{i \gamma T_8} 
    V_{\mathrm{CKM}} 
    e^{i \beta^\prime T_3} e^{i \gamma^\prime T_8}. 
\label{phases}
\ee
However, in this paper we adopt these phases 
$\alpha$, $\beta$, $\gamma$, $\beta^\prime$ and $\gamma^\prime$ 
to zero or $\pi$.  So $M_{10}$ and $M_{126}$ are the functions 
of $\sigma$, the phase of $c_d$, with the solutions, 
$|c_d|$ and $\kappa$, determined by the GUT relation.  

\begin{figure}[p]
\begin{center}
\includegraphics{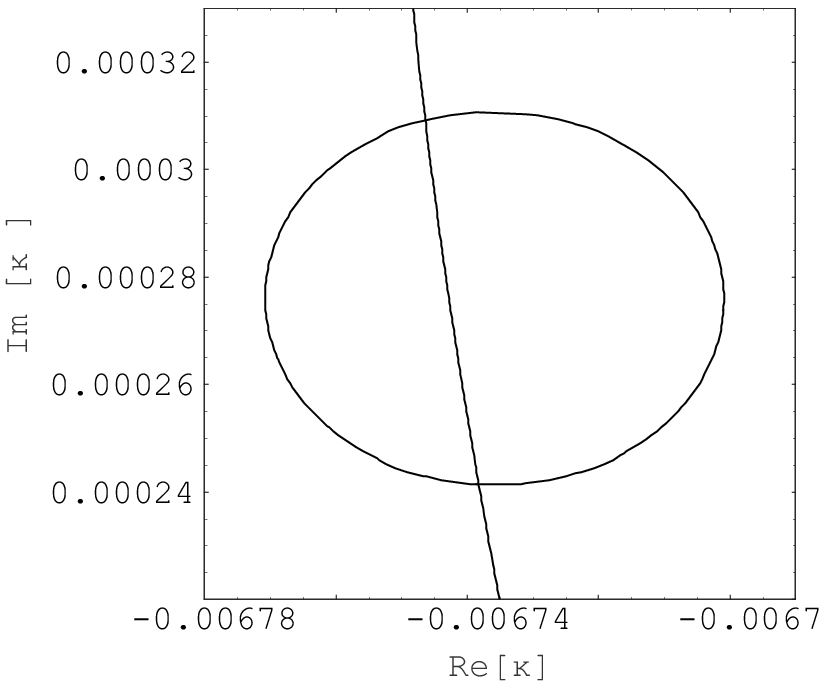}
\end{center}
\caption{Contour plot on complex $\kappa$-plane in case 
 of $\tan\beta = 2.5$.  
The vertical line and the circle correspond to the solutions 
 of Eqs.~(\ref{cond1})  and (\ref{cond2}), respectively.  }
\label{f1}
\end{figure}

Now let us solve the GUT relation and determine $|c_d|$ and $\kappa$. 
Since the GUT relation of Eq.~(\ref{GUTrelation}) is valid 
 only at the GUT scale, 
 we first evolve the data at the weak scale 
 to the ones at the GUT scale with given $\tan \beta$ 
 according to the renormalization group equations (RGEs)
 and use them as input data at the GUT scale. 
Note that it is non-trivial to find the solution 
 of the GUT relation, 
 since the number of the free parameters (fourteen) is 
 almost the same as the number of inputs (thirteen).  
The solution of the GUT relation exists, 
 only if we take appropriate input parameters.  
Therefore, in the following analysis, 
 we vary two input parameters, $m_s$ and CP-phase $\delta$ 
 in the CKM matrix, within the experimental errors 
 so as to find the solution.  
We take input fermion masses at $M_Z$ as follows (in GeV): 
\begin{eqnarray}
& & m_u = 0.00233 \; , \; \;   m_c = 0.677 \; , \; \;  m_t=176, 
\nonumber\\
& & m_d = 0.00469 \; , \; \;  m_b = 3.00, 
\nonumber\\ 
& & m_e = 0.000487 \; , \; \; m_\mu=0.103 \; ,
\; \;  m_\tau=1.75. 
\nonumber 
\end{eqnarray} 
Here the experimental values extrapolated from low energies 
 to $M_Z$ were used \cite{fk}, and we choose the signs of 
 the input fermion masses as $-$ for $m_c$, $m_d$ and $m_s$, 
 and $+$ for $m_u$, $m_t$ and $m_b$.  
For the CKM mixing angles in the ``standard'' parameterization, 
 we input the center values measured by experiments 
 as follows \cite{PDG}: 
\begin{eqnarray}
 s_{12} = \frac{0.219+0.226}{2} , \; \;  
 s_{23} = \frac{0.037+0.043}{2} , \; \;
 s_{13} = \frac{0.002+0.005}{2} \; . \nonumber 
\end{eqnarray} 
In the following, we show our analysis in detail in two cases 
 $\tan \beta=2.5$ and $10$.  The lower bound on $\tan \beta 
> 2.4$ comes from the LEP Higgs searches \cite{LEP}.  
We take $m_s = 70.1 \,\mathrm{[MeV]}$ at $M_Z$ scale
 for both cases of $\tan\beta = 2.5$ and $10$, 
 but take $\delta= 132^\circ$ 
 and $\delta= 108^\circ$ for $\tan\beta = 2.5$ and $10$, respectively.  
After RGE evolutions to the GUT scale, fermion masses (up to sign) 
 and the CKM matrix are found as follows (in GeV): 
For $\tan\beta =2.5$, 
\begin{eqnarray}
& & m_u = 0.00126 \; , \; \; m_c = 0.365 \; , \; \;  m_t = 184, 
\nonumber\\
& & m_d = 0.00134 \; , \; \;  m_s = 0.0200 \; , \; \; m_b = 1.07, 
\nonumber\\ 
& & m_e = 0.000323 \; , \; \; m_\mu = 0.0682 \; , \; \;  m_\tau = 1.16,  
\nonumber
\end{eqnarray} 
and 
\begin{eqnarray}
V_{\mathrm{CKM}}(M_G) 
= \left( 
\begin{array}{ccc}
0.975 & 0.222 & - 0.00188 - 0.00208 i \\
- 0.222 - 0.000101 i & 0.974 + 0.000142 i & 0.0321 \\ 
0.00896 - 0.00203 i & - 0.0308 - 0.000468 i & 0.999
\end{array} 
\right) \;   ,
\nonumber
\end{eqnarray}
and for $\tan\beta =10$, 
\begin{eqnarray}
& & m_u = 0.000980 \; , \; \; m_c = 0.285 \; , \; \;  m_t = 113, 
\nonumber\\
& & m_d = 0.00135 \; , \; \; m_s = 0.0201 \; , \; \; m_b = 0.996, 
\nonumber\\ 
& & m_e = 0.000326 \; , \; \; m_\mu = 0.0687 \; , \; \; m_\tau = 1.17, 
\nonumber
\end{eqnarray} 
and 
\begin{eqnarray}
V_{\mathrm{CKM}}(M_G) 
= \left( 
\begin{array}{ccc}
0.975 & 0.222 & - 0.000940 - 0.00289 i \\
-0.222 - 0.000129 i & 0.974 + 0.000124 i & 0.0347 \\ 
0.00864 - 0.00282 i & - 0.0337 - 0.000647 i & 0.999
\end{array} 
\right) \; .  
\nonumber
\end{eqnarray}
in the standard parameterization.  
By using these outputs at the GUT scale as input parameters, 
 we solve Eqs.~(\ref{cond1}) and (\ref{cond2}). 
The contours of solutions of each equations are depicted 
 in Figure \ref{f1} for $\tan \beta = 2.5$. 
The crossing points of two contours are the solutions. 
As an example, we list a solution (at the upper crossing point) 
\begin{eqnarray}
& \kappa =
-0.00675 + 0.000309 i \;,   \nonumber\\
& |c_d| = 5.99  \; , & 
\end{eqnarray}
for $\tan\beta = 2.5$, and 
\begin{eqnarray}
& \kappa = - 0.0103 + 0.000606 i \;, \nonumber\\ 
& |c_d| = 6.32  \; , & 
\end{eqnarray}
for $\tan\beta = 10$.  

Once the parameters, $|c_d|$ and $\kappa$, are determined, 
 we can describe all the fermion mass matrices as a functions 
 of $\sigma$ from the mass matrix formulas 
 of Eqs.~(\ref{massmatrix}), (\ref{M10}) and (\ref{M126}).  
Interestingly, in the minimal $SO(10)$ model
 even light Majorana neutrino mass matrix, $M_{\nu}$, 
 can be determined as a function of $\sigma$, $c_T$ and $c_R$ 
 through the seesaw mechanism 
 $M_{\nu}= - M_D^T M_R^{-1} M_D + M_T$ . 
The case where the the first term dominates 
 is called type I seesaw model, 
 while the case where the the second term dominates 
 is called type II seesaw model. 
Each cases have been analyzed in detail 
 in \cite{fo} and \cite{gmn}, respectively, 
 and discussed the consistency with 
 the current neutrino oscillation data. 
Recently the general case has been analyzed, 
 it is found that 
 the minimal $SO(10)$ model is not so good 
 to fit all the neutrino oscillation data \cite{mimura}. 
This means that the structure of the model is 
 somewhat too restrictive for the neutrino sector. 
It would be inevitable to extend the model 
 at least for the neutrino sector 
 in order to make the data fitting much better. 
Note that there are lots of possible ways 
 to minimally extend the model only for the neutrino sector
 but keep the predictive power for the charged fermion mass matrices. 
Remember that there are two free parameters relevant to 
 the neutrino sector, $v_T$ and $v_R$, in the model. 
If $v_T$ is smaller than the mass scale 
 of the neutrino oscillation data, 
 and if $v_R$ is the GUT scale%
\footnote{
This is natural if the $SO(10)$ group is broken down to 
 the standard model one at the GUT scale. 
}
the resultant light neutrino mass eigenvalues 
 through the seesaw mechanism are too small 
 to be compatible with the scale of the neutrino oscillation data. 
In such a case, we have to extend the neutrino sector. 
For instance, we can introduce an additional 
 ${\bf \overline{126}}$ multiplet 
 which admits to obtain the VEV only in the $SU(2)_L$ triplet 
 direction, $v_T^\prime$. 
Now suppose that the type II seesaw mechanism works 
 and we obtain the light neutrino Majorana mass matrix 
 as $Y_{126}^\prime v_T^\prime$, 
 where $Y_{126}^\prime$ is newly introduced Yukawa coupling. 
If each elements in $Y_{126}^\prime$ are much smaller than 
 that in $Y_{126}$, 
 our analysis above remains correct
 and the predictive power for the charged fermion mass matrices 
 is maintained. 
In the following analysis, we assume 
 such a minimal extension of the model.

\section{Proton decay via dimension five operator} 
%
The Yukawa interactions of the MSSM matter with 
 the color triplet Higgs induces the following 
 Baryon and Lepton number violating dimension five operator 
\be
W=C_L^{ijkl} Q^i Q^j Q^k L^l. 
 \label{dim5}
\ee
Here the coefficients are given by the products of 
 the Yukawa coupling matrices and 
 the (effective) color triplet Higgsino mass matrix, 
 and are model dependent.  
In the minimal $SO(10)$ model, 
 the coefficients are given by the products of 
 two basic Yukawa coupling matrices, 
 $Y_{10}$ and $Y_{126}$, 
 and 
 the effective $2 \times 2$ color triplet 
 Higgsino mass matrix, $ M_C $, such as  \cite{fikmo} 
\be
C_L^{ijkl} 
= 
\left(Y_{10}^{ij},~Y_{126}^{ij} \right)
\left( M_C^{-1}  \right)
\left( 
\begin{array}{c}
  Y_{10}^{k l} \\
  Y_{126}^{k l}
\end{array}  \right) .
\label{CL}
\ee

As discussed in the previous section 
 the Yukawa coupling matrices, $Y_{10}$ and $Y_{126}$, 
 are related to the corresponding mass matrices 
 $M_{10}$ and $M_{126}$ such that 
\bea
Y_{10} &=& \frac{c_{10}}{\alpha^u  v \sin{\beta}} M_{10}, 
\nonumber\\
Y_{126} &=& \frac{c_{126}}{\beta^u  v \sin{\beta}} M_{126}, 
\eea
with $v \simeq 174.1 \, \mathrm{[GeV]}$.  
Here $\alpha^u$ and $\beta^u$ are the Higgs doublet mixing parameters
 introduced in the previous section, 
 which are restricted in the range 
 $|\alpha^u|^2 +|\beta^u|^2 \leq 1$. 
Although these parameters are irrelevant 
 to fit the low energy experimental data 
 of the fermion mass matrices, 
 there are theoretical lower bound on them 
 in order for the resultant Yukawa coupling constant 
 not to exceed the perturbative regime. 
Since $c_{10}$, $c_{126}$, $M_{10}$ and $M_{126}$ 
 are the functions of only $\sigma$, 
 we can completely determine the Yukawa coupling matrices 
 once $\sigma$, $\alpha^u$ and $\beta^u$ are fixed. 
In order to obtain the most conservative values of 
 the proton decay rate, 
 we make a choice of the Yukawa coupling matrices 
 as small as possible. 
In the following analysis, we restrict the region 
 of the parameters in the range 
 $(\alpha^u)^2 +(\beta^u)^2 = 1$ 
 (we assume $\alpha^u$ and $\beta^u$ real for simplicity). 
Here we present examples of the Yukawa coupling matrices 
 with fixed $\sigma = \pi$.  
For $\tan\beta = 2.5 $ with $\alpha^u = 0.031$, we find  
\be
Y_{10}
=
\left(
\begin{array}{ccc}
\begin{array}{c}
0.000839 + 2.79 \times {10}^{-6} \,i \\
0.00151 - 0.0000265 \, i \\
0.000692 - 0.000818 \, i \\
\end{array}
\begin{array}{c}
0.00151 - 0.0000265 \, i \\
0.00479 + 0.0000811 \, i \\
-0.0128 + 3.17 \times {10}^{-6} \, i 
\end{array}
\begin{array}{c}
0.000692 - 0.000818 \, i \\
-0.0128 + 3.17 \times {10}^{-6} \, i \\
0.525 - 0.0420 \, i 
\end{array}
\end{array}
\right),
\ee
\be
Y_{126}
=
\left(
\begin{array}{ccc}
\begin{array}{c}
-0.0000613 - 2.17 \times {10}^{-7} \, i \\
-0.000111 + 1.94 \times {10}^{-6} \, i \\
-0.0000508 + 0.0000600 \, i 
\end{array}
\begin{array}{c}
-0.000111 + 1.94 \times {10}^{-6} \, i \\
-0.000428 - 2.46 \times {10}^{-6} \, i \\
0.000941 - 2.33 \times {10}^{-7} \, i 
\end{array}
\begin{array}{c}
-0.0000508 + 0.0000600 \, i \\
0.000941 - 2.33 \times {10}^{-7} \, i \\
1.42 \times {10}^{-7} + 0.00132 \, i
\end{array}
\end{array}
\right),  
\ee
and for $\tan\beta = 10$ with $\alpha^u = 0.111$,  
\be
Y_{10}
=
\left(
\begin{array}{ccc}
\begin{array}{c}
0.00101 + 1.87 \times {10}^{-6} \, i \\
0.00179 - 0.0000439 \, i \\
0.000348 - 0.00125 \, i 
\end{array}
\begin{array}{c}
0.00179 - 0.0000439 \, i \\
0.00541 + 0.000141 \, i \\
-0.0154 + 5.21 \times {10}^{-6} \, i 
\end{array}
\begin{array}{c}
0.000348 - 0.00125 \, i \\
-0.0154 + 5.21 \times {10}^{-6} \, i \\
0.530 - 0.0567 \, i 
\end{array}
\end{array}
\right),
\ee
\be
Y_{126}
=
\left(
\begin{array}{ccc}
\begin{array}{c}
-0.000244 - 5.21 \times {10}^{-7} \, i \\
-0.000436 + 0.0000107 \, i \\
-0.0000847 + 0.000305 \, i 
\end{array}
\begin{array}{c}
-0.000436 + 0.0000107 \, i \\
-0.00164 - 0.0000154 \, i \\
0.00375 - 1.27 \times {10}^{-6} \, i 
\end{array}
\begin{array}{c}
-0.0000847 + 0.000305 \, i \\
0.00375 - 1.27 \times {10}^{-6} \, i \\
-0.000684 + 0.00626 \, i 
\end{array}
\end{array}
\right).  
\ee

For the effective color triplet Higgsino mass matrix, 
 we assume the eigenvalues being the GUT scale, 
 $M_G = 2 \times 10^{16} \, \mathrm{[GeV]}$, 
 which is necessary to keep the successful gauge 
 coupling unification. 
Then, in general, we can parameterize the $2 \times 2$ 
 mass matrix as 
\be
M_C  = M_G I_2 \times U, 
\label{MT}
\ee
 with the unitary matrix, 
\be
U = e^{i \varphi \sigma_3} 
 \left(
 \begin{array}{cc}
 \begin{array}{c}
\cos{\theta} \\
- \sin{\theta}
\end{array}
\begin{array}{c}
\sin{\theta} \\
\cos{\theta}
\end{array} 
\end{array}
  \right) 
e^{i \varphi^\prime \sigma_3}.  
\ee
Here we omit an over all phase since it is irrelevant 
 to calculations of the proton decay rate.  
Now there are five free parameters in total 
 involved in the coefficient $C_L^{ i j k l}$, 
 namely, $\sigma$, $\alpha_u$, $\theta$, 
 $\varphi$ and $\varphi^\prime$. 
Once these parameters are fixed, $C_L^{i j k l}$ 
 is completely determined. 

The proton decay mode via the dimension five operator 
 in Eq.~(\ref{dim5}) 
 with the Wino dressing diagram 
 is found to be dominant, and leads to 
 the proton decay process, $p \to K^{+} \overline{\nu}$. 
The decay rate for this process 
 is approximately estimated as 
 (in the leading order of the Cabibbo angle $\lambda \sim 0.22$) 
\bea
\Gamma(p \to K^{+} \overline{\nu})
& \simeq &
\Gamma(p \to K^{+} \overline{\nu_\tau})
=
\frac{m_p}{32 \pi f_{\pi}^2} \,
\left|\beta_{H} \right|^2 \times
\left|A_L A_S \right|^2 \times
\left(\frac{\alpha_2}{4 \pi} \right)^2 
\frac{1}{m_{S}^2}
\nonumber\\
&\times&
 \Big|C_L^{2311} - C_L^{1312} 
  + \lambda \left( C_L^{2312} - C_L^{1322} \right)\Big|^2 
\nonumber\\
&\times& 
 5.0 \times 10^{31} \, [{\mathrm{years^{-1}/GeV}}].  
\label{rate1}
\eea
Here the first term denotes 
 the phase factor and the hadronic factor, 
 $\beta_{H} = 0.0096 \,[{\mathrm GeV^3}]$ 
 given by lattice calculations \cite{lattice}. 
$A_L \simeq 0.32$, $A_S \simeq 0.93$ are 
 the long-distance and the short-distance 
 renormalization factors about the coefficient $C_L^{ijkl}$, 
 respectively.  
\footnote{As suggested in Ref. \cite{turzynski}, 
it might be proper to use the renormalization factors 
$\tilde{A}_L$, $\tilde{A}_S$ in \cite{turzynski}, which directly treats 
the renormalization of the Wilson coefficients itself.  
But here, we adopt the use of the conventional factors $A_L$, $A_S$ 
to compare our results to the previous ones.  }
The third term in the first line comes from 
 the Wino dressing diagram, and 
 $m_S$ is a typical sparticle mass scale 
 multiplied by the ratio of a sfermion and Wino. 
In the case with the mass hierarchy 
 between the sfermions and the Wino 
 $(\widetilde{m}_f \gg M_2)$, 
 we find $m_S \sim \widetilde{m}_f \times ({\widetilde{m}_f}/M_2)$. 
In the following numerical analysis, 
 we take $\widetilde{m}_f  = 1 \,{\mathrm{[TeV]}}$ 
 and $M_2  = 100 \,{\mathrm{[GeV]}}$. 

\begin{figure}[p]
\begin{center}
\includegraphics{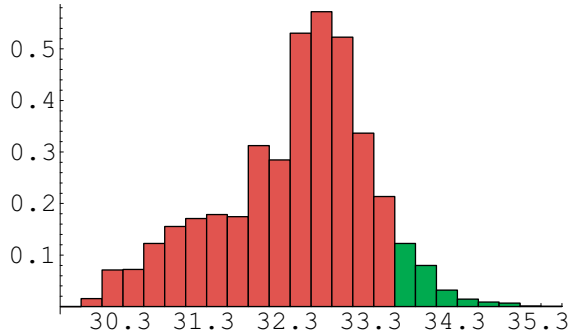}
\end{center}
\caption{
The distributions of the proton lifetime (log years) 
 for $\tan \beta =2.5$ 
 in arbitrary five parameter choices (normalized by 1).  
The green region (right hand side) is consistent with experiment. 
}
\label{f3}
\end{figure}

\begin{figure}[p]
\begin{center}
\includegraphics{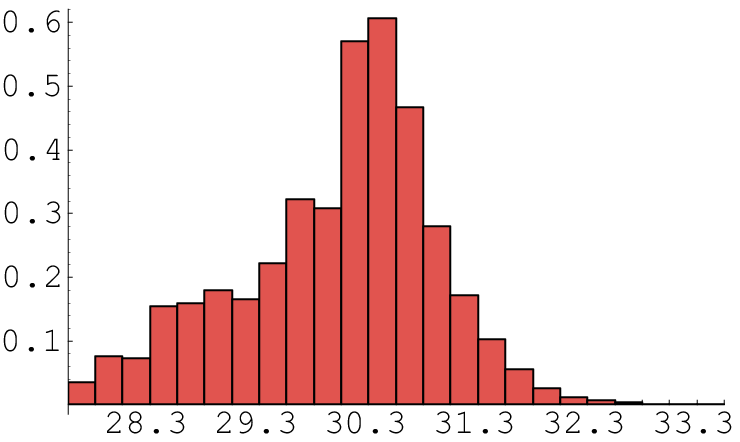}
\end{center}
\caption{Same as in Figure \ref{f3}, but for $\tan\beta=10$.  }
\label{f4}
\end{figure}

Now we perform numerical analysis. 
Note that because of the very constrained flavor structure 
 of the minimal SUSY $SO(10)$ model 
 we can give definite predictions for the proton decay rate 
 once the five parameters in the above are fixed. 
For a specific choice of the Yukawa coupling matrices 
 in the minimal $SO(10)$ model with the type II seesaw, 
 the proton decay rate has been calculated in \cite{gmnn}. 
In our analysis, we make no such a specific choice, 
 and perform detailed analysis 
 in general situations of the minimal $SO(10)$ model
 by varying the above five free parameters. 
The result for $\tan \beta =2.5$ is presented in Figure \ref{f3}. 
Here the distributions of the proton lifetime (log years) 
 for arbitrary choices of the five free parameters (normalized by 1) 
 is depicted. 
We can see that some special sets of the free parameters 
 can result the proton lifetime consistent with SuperK results. 
In that region, cancellation in the second line 
 in Eq.~(\ref{rate1}) occurs 
 by tuning of the free parameters in the Higgsino mass matrix. 
Note that number of free parameters is not enough 
 to cancel both of the process 
 $p \to K^{+} \overline{\nu_\tau}$ (dominant mode) and 
 $p \to K^{+} \overline{\nu_\mu}$ (sub-dominant mode), 
 and thus the proton lifetime has an upper bound 
 in the $SO(10)$ model. 
For $\tan\beta = 10$, we obtain the same figure 
 depicted in Figure \ref{f4} 
 but the lifetime is scaled by roughly $(2.5/10)^2$, 
 which is consistent with the naive expectation 
 that the lifetime is proportional to $1/\tan^2 \beta$. 
Whole region is excluded in the case with $\tan \beta=10$. 

In the case of nondegenerate masses of $M_C$, 
the parameters increase from three to $5+1$ 
(the last 1 is the ratio of masses). 
However, the results only slide by the square of 
this mass ratio and do not show the special cancellation. 

\section{Conclusion and discussions}

We have discussed the minimal SUSY SO(10) model, 
 which can reproduce the realistic charged fermion mass matrices 
 with only one parameter left free. 
This model has high predictive power for 
 the fermion Yukawa coupling matrices, 
 and they are completely determined 
 once a few parameters in the model fixed.
This feature is essential for definite predictions 
 of the proton decay rate vis the dimension five operators. 
Including additional 3 free parameters 
 in the effective Higgsino mass matrix 
 with mass eigenvalue being the GUT scale, 
 we have analyzed the proton decay rate 
 by varying the five free parameters in total. 
We have found that for $\tan \beta=2.5$ 
 some special sets of the parameters 
 predicts the proton decay rate consistent with the SuperK results, 
 where the cancellation for the dominant modes 
 of the proton decay amplitude occurs 
 by tuning of the parameters. 
Although there exists the allowed region, it is very narrow. 
Our result is consistent with the one in the previous work \cite{gmnn} 
 for only one specific choice of the Yukawa coupling matrices. 
It has been found that the resultant proton decay rate 
 is proportional to $\tan^2 \beta$ as expected 
 and the allowed region eventually disappears 
 as $\tan \beta$ becomes large, even for $\tan \beta=10$. 

There are some theoretically possible ways to 
 extend the proton lifetime. 
One way is to adopt a large mass hierarchy 
 between the sfermions and the Wino 
 as can be seen in Eq.~(\ref{rate1}). 
The proton lifetime is pushed up 
 according to the squared powers of the mass hierarchy, 
 and the allowed region becomes wide. 
How large the hierarchy can be depends on 
 the mechanism of the SUSY breaking and its mediation. 
When we assume the minimal supergravity scenario, 
 the cosmologically allowed region \cite{CMSSM} 
 consistent with the recent WMAP satellite data \cite{WMAP} 
 suggests that masses between sfermion masses the Wino 
 is not so hierarchical and the value we have taken 
 in our analysis seems to be reasonable. 
Another way is to abandon the assumption 
 of Higgsino degeneracy at the GUT scale, 
 and to make the mass eigenvalues of 
 the effective colored Higgsino mass matrix heavy. 
We can examine this possibility based 
 on a concrete Higgs sector. 
However this seems to be a very difficult task 
 even if we introduce a minimal Higgs sector 
 in the minimal $SO(10)$ model discussed 
 in \cite{fikmo} \cite{bmsv}, 
 since there are lots of free parameters in the Higgs sector. 
Furthermore, even in the minimal Higgs sector, 
 there are lots of Higgs multiplets involved 
 and the beta function coefficients of the gauge couplings are huge. 
It seems to be very hard 
 to succeed the gauge coupling unification 
 before blowing up of the gauge couplings. 
Therefore, the assumption that all the Higgs multiplets 
 are degenerate at the GUT scale would be natural.  
Consequently our results show the typical properties of 
SO(10) GUT but are not exhaustive. Also there is possibility to 
vary GUT phases $\alpha$, $\beta$, $\gamma$, $\beta^\prime$ and 
$\gamma^\prime$ in Eq. (\ref{phases}).

\section*{Acknowledgment}
The work of A.I. and S.M. is supported by the Ministry of the Science 
 and Technology of the Republic of Croatia. 
The work of T.F. , T.K. and N.O. is supported 
 by the Grant in Aid for Scientific Research 
 from the Ministry of Education, Science and Culture. 
The work of T.K. is also supported by the Research Fellowship 
 of the Japan Society for the Promotion of Science 
 for Young Scientists. 
T.F. would like to thank G. Senjanovic and A.Y. Smirnov 
 for their hospitality at ICTP.  
T.K. is grateful to K. Turzynski for his useful comments and discussions 
about renormalization procedures.  
Also T.F. and T.K. are grateful to K. Matsuda 
 for his useful comments on numerical analysis.
N.O. would like to thank Y. Mimura and D. Chang 
 for useful discussions. 

%
%
%

%
%
\newpage

\begin{thebibliography}{99}
%
\bibitem{gut}
H. Georgi and S.L. Glashow, 
 Phys. Rev. Lett. {\bf 32}, 438 (1974). 
%
\bibitem{susygut2}
U. Amaldi, W.de Boer, and H. F{\"u}rstenau, 
 Phys. Lett. {\bf B260}, 447 (1991); 
P. Langacker and M. Luo, 
 Phys. Rev. {\bf D44}, 817 (1991).
%
\bibitem{sk}
M. Shiozawa, 
talk at the 4th workshop on 
"Neutrino Oscillations and their Origin" (NOON 2003), 
[http://www-sk.icrr.u-tokyo.ac.jp/noon2003/].  
%
\bibitem{mp}
H. Murayama and A. Pierce, 
 Phys. Rev. {\bf D65}, 055009 (2002).  
%
\bibitem{gn}
T. Goto and T. Nihei, Phys. Rev. {\bf D59}, 115009 (1999).  
%
\bibitem{bps}
B. Bajc, P.F. Perez and G. Senjanovic, 
Phys. Rev. {\bf D66}, 075005 (2002).  
%
\bibitem{ew}
D. Emmanuel-Costa and S. Wiesenfeldt,
Nucl. Phys. B {\bf 661}, 62 (2003)
%
\bibitem{seesaw}
T. Yanagida, in Proceedings of the workshop 
on the Unified Theory and Baryon Number in the Universe, 
edited by O. Sawada and A. Sugamoto (KEK, Tsukuba, 1979);
M. Gell-Mann, P. Ramond, and R. Slansky, 
in Supergravity, edited by D. Freedman and P. van Nieuwenhuizen 
(North-Holland, Amsterdam, 1979); 
R.N. Mohapatra and G. Senjanovic, 
Phys.\ Rev.\ Lett.\  {\bf 44}, 912 (1980).  
%
\bibitem{bm}
K.S. Babu and R.N. Mohapatra, 
Phys. Rev. Lett. {\bf 70}, 2845 (1993).  
%
\bibitem{fo}
T. Fukuyama and N. Okada, 
 JHEP {\bf 0211}, 011 (2002); 
K. Matsuda, Y. Koide, T. Fukuyama and H. Nishiura,
 Phys. Rev. {\bf D65}, 033008 (2002) 
 [Erratum-ibid. {\bf D65}, 079904 (2002)]; 
K. Matsuda, Y. Koide and T. Fukuyama,
 Phys. Rev. {\bf D64}, 053015 (2001).  
%
\bibitem{gmn}  
B. Bajc, G. Senjanovic and F. Vissani, 
 Phys. Rev. Lett. {\bf 90}, 051802 (2003); 
H.S. Goh, R.N. Mohapatra and Siew-Phang Ng, 
 Phys. Lett. {\bf B570}, 215 (2003),  
 Phys. Rev. {\bf D68}, 115008 (2003).  
%
\bibitem{mimura}
B. Dutta, Y. Mimura and R.N. Mohapatra, 
 arXiv: hep-ph/0402113. 
%
\bibitem{perez}
P.F. Perez, arXiv: hep-ph/0403286. 
%
\bibitem{ps}
J.C. Pati and A. Salam, Phys. Rev. {\bf D10}, 275 (1974). 
%
%
\bibitem{fk}
H. Fusaoka and Y. Koide, 
 Phys. Rev. {\bf D57}, 3986 (1998). 
\bibitem{PDG}
K. Hagiwara {\it et al.} [Particle Data Group Collaboration], 
 Phys. Rev. {\bf D66}, 010001 (2002).  
%
\bibitem{LEP}
LEP Higgs Working Group and ALEPH collaboration and DELPHI 
collaboration and L3 collaboration and OPAL Collaboration, 
arXiv: hep-ex/0107030.  
%
\bibitem{fikmo}
T. Fukuyama, A. Ilakovac, T. Kikuchi, S. Meljanac, and N. Okada, 
arXiv: hep-ph/0401213.  
%
\bibitem{lattice}
N. Tsutsui {\it et al.} [CP-PACS collabolation and JLQCD collaboration], 
[arXiv:hep-lat/0402026].  
%
\bibitem{turzynski}
K. Turzynski, JHEP {\bf 0210}, 044 (2002). 
%
\bibitem{gmnn}
H.S. Goh, R.N. Mohapatra, S. Nasri, Siew-Phang Ng, 
Phys. Lett. {\bf B587}, 105 (2004).  
%
\bibitem{CMSSM}
For a review article, for example, 
A.B. Lahanas, N.E. Mavromatos, D.V. Nanopoulos, 
Int. J. Mod. Phys. {\bf D12} 1529, (2003).  
%
\bibitem{WMAP}
C.L. Bennett {\it et al.}, Astrophys. J. Suppl. {\bf 148}, 1 (2003); 
D.N. Spergel {\it et al.}, Astrophys. J. Suppl. {\bf 148}, 175 (2003).  
%
\bibitem{bmsv}
B. Bajc, A. Melfo, G. Senjanovic and F. Vissani,
 arXiv:hep-ph/0402122.
%
%
\end{thebibliography}
\end{document}